\documentclass[12pt]{article}
\usepackage{a4wide}
\usepackage{amssymb}

\begin{document}

{\renewcommand{\thefootnote}{\fnsymbol{footnote}} 
\medskip }

\begin{center}
{\LARGE Quantum matter in quantum space-time}\\[0pt]
\vspace{1.5em} Martin Bojowald,$^1$ 
Golam Mortuza Hossain,$^2$ Mikhail Kagan$^{1,3}$ and Casey Tomlin$^{1,4}$ 
\\[0pt]
\vspace{0.5em} $^1$Institute for Gravitation and the Cosmos,\\[0pt]
The Pennsylvania State University,\\[0pt]
104 Davey Lab, University Park, PA 16802, USA\\[0pt]
\vspace{0.5em} $^2$Department of Physical Sciences, \\[0pt]
Indian Institute of Science Education and Research Kolkata, \\[0pt]
Mohanpur Campus, PO: Krishi Viswavidyalaya, Nadia - 741 252, WB, India 
\\[0pt]
\vspace{0.5em} $^3$Department of Science and Engineering, \\[0pt]
The Pennsylvania State University, Abington\\[0pt]
1600 Woodland Road, Abington, PA 19116, USA\\[0pt]
\vspace{0.5em} $^4$Max Planck Institute for Gravitational Physics (Albert
Einstein Institute),\\[0pt]
Am M\"uhlenberg 1, D-14476 Golm, Germany\\[0pt]
\vspace{1.5em}
\end{center}

\begin{abstract}
Quantum matter in quantum space-time is discussed using general properties
of energy-conservation laws. As a rather radical conclusion, it is found
that standard methods of differential geometry and quantum field theory on
curved space-time are inapplicable in canonical quantum gravity, even at the
level of effective equations.
\end{abstract}

\setcounter{footnote}{0}

\section{Introduction}

Energy conservation is the most important general statement about matter,
classical and quantum. Its fundamental role is strengthened by its relation
to space-time symmetries. As expressed by Hamiltonian equations in classical
physics or, even more directly, by the operator relationships $\hat{\vec{p}}%
=-i\hbar\vec{\nabla}$ and $\hat{E}=i\hbar\partial/\partial t$ of quantum
mechanics, momentum generates spatial shifts and energy generates time
translations. When local densities are used as energy and momentum
expressions in field theory, they are related to local coordinate
transformations in space-time instead of rigid global shifts. Energy
conservation is therefore closely related to general covariance, the
underlying symmetry principle of space-time.

Matter fields are quantized in quantum field theory, in which operators take
the form of energy and momentum densities and are still conserved. The close
relation between energy-momentum and space-time symmetries suggests that
quantum corrections in the former might affect even the form of general
covariance and therefore the structure of space-time. This expectation is
not borne out in quantum field theory on curved space-time because of the
conservation law: for operators, it takes the same form as for classical
densities, with covariant derivatives acting on the dependence of quantum
fields on (classical) coordinates. In quantum gravity, however, space-time
is quantized and the classical notions of differentiable manifolds and
coordinates may lose their meaning or become inapplicable. Reversing the
preceding argument, it is then conceivable that not only space-time
structure but also the corresponding form of energy conservation changes.

Quantum gravity is still being constructed and complicated to use. The link
between space-time and energy-momentum of matter then offers the possibility
of easier insights using matter alone, but on a background endowed with
expected features of quantum space-time. In this article, we begin with
existing results about deformed covariance principles in canonical quantum
gravity, inserted in the energy conservation law, and aim to draw
conclusions about possible structures of differential geometry in quantum
space-time. One could expect that the covariant derivative prominent in
conservation laws on curved space-time would have to be modified for
quantum-gravity effects. Such an implication would be of interest for an
intuitive understanding of the underlying quantum geometry, for instance in
terms of possible relationships with non-commutative \cite{Connes,NonCommST}
or fractional \cite{Fractional} models.\footnote{%
For additional consequences of deformed general covariance,  see \cite%
{Action,DeformedRel}.} Moreover, a direct modification of the conservation
law of matter would be of great interest for cosmological perturbation
theory, whose equations in terms of gauge-invariant quantities can be
derived from the behavior of stress-energy without reference to the more
complicated Einstein equation \cite{NonLinPert}.\footnote{%
We thank  Gianluca Calcagni for stressing this.}

To embark on these investigations, in the main body of this article we
review canonical gravity in terms of transformations between different
families of observers in space-time and rewrite the usual covariant
conservation law in canonical terms. These details will show how space-time
symmetries play a key role for the validity of energy conservation, and how
possible deformations of symmetries by quantum effects could change
conservation laws. Our conclusion is rather radical: It is not possible to
modify the conservation law by mere coefficients of derivative and
connection terms in its space-time form; rather, some quantum theories of
gravity indicate that the usual space-time tensor calculus breaks down even
at the level of effective theories, while canonical methods remain
consistent. The latter appear to be more fundamental than action principles,
providing concrete evidence for a long-standing claim by Dirac \cite%
{DiracHamGR}. Moreover, standard techniques of quantum field theory on
curved space-time cannot be used because they are closely tied to
conventional conservation laws.

\section{Energy conservation}

In Minkowski space-time, energy conservation can compactly be written as 
\begin{equation}  \label{dT}
\partial_{\mu} T^{\mu}{}_{\nu}=0
\end{equation}
with the stress-energy tensor $T_{\mu\nu}$ containing the energy density as
its time-time component, momentum density and energy flux as the mixed
time-space components, and spatial stress and pressure in its spatial part.
Stokes' theorem, applied to (\ref{dT}) integrated over a region in
space-time, then shows that the change of energy in spatial cross-sections
equals the energy flux through timelike boundaries.

The relation to space-time symmetries becomes apparent when one extends the
conservation law to matter in curved space-time, with metric tensor $%
g_{\mu\nu}$. Lorentz transformations of Minkowski space-time are replaced by
local coordinate transformations, invariance under which implies the
conservation law 
\begin{equation}  \label{DT}
\nabla_{\mu} T^{\mu}{}_{\nu}=\partial_{\mu}T^{\mu}{}_{\nu}+
\Gamma^{\mu}_{\mu\rho} T^{\rho}{}_{\nu}- \Gamma^{\rho}_{\mu\nu}
T^{\mu}{}_{\rho}=0
\end{equation}
with a covariant derivative instead of the partial one. The link between
energy-momentum and the space-time metric also appears in the equation 
\begin{equation}  \label{T}
T_{\mu\nu}= \frac{-2}{\sqrt{|\det g|}}\frac{\delta S_{\mathrm{matter}}}{%
\delta g^{\mu\nu}}
\end{equation}
with the matter action $S_{\mathrm{matter}}$.

In general relativity, the metric encodes the gravitational field, and it is
quantized in quantum gravity. One could worry that quantum corrections in
the metric might modify or even violate energy conservation. As one may
expect for such basic laws, however, they are protected by general
principles that do not refer to details of the form of matter. The link
between energy conservation and space-time symmetries ensures that any
theory, classical or quantum, that is independent of choices of coordinates
gives rise to energy conservation. The independence of choices of
coordinates is not just a key feature of classical general relativity but
also an important consistency requirement for quantum gravity. Systems of
coordinates are, after all, mere choices to set up mathematical
descriptions, which cannot affect physical predictions. Quantum gravity must
enjoy the same degree of invariance, or it is not consistent as a physical
theory. Even though it is complicated to ensure this invariance or
anomaly-freedom, posing perhaps the main obstacle to a successful completion
of quantum gravity, the generality of the requirement allows us to use it
for conclusions about energy conservation.

\section{Canonical gravity}

Canonical gravity provides powerful methods to analyze space-time structure.
Its quantum branch, canonical quantum gravity, has given rise to the most
detailed results about possible quantum corrections and deformations in
covariance laws. It provides the main ingredients of quantum space-time
structure we will use in this article. To this end, we first recall crucial
features of the canonical theory worked out starting with \cite{DiracHamGR, ADM}; for a detailed treatment see \cite{CUP}.

In canonical gravity, one describes covariant space-time as seen by families
of observers who have undertaken different synchronizations of their clocks.
For each family, there is a notion of the observers' proper time which, as
the set of all points taking a given fixed value, determines spatial
cross-sections in space-time. A single cross-section amounts to space at an
instance of time according to the family of observers used, but a different
family with its own synchronization will see different cross-sections of
space-time as space. Canonical gravity provides laws to transform between
the viewpoints of different families, as local generalizations of Lorentz
transformations in Minkowski space-time.

A given family of observers moves in space-time along worldlines with
4-velocity equal to the future-pointing unit normal $u^{\mu}=n^{\mu}$ of
spatial cross-sections according to its proper time. In each cross-section,
distances are measured with a spatial metric tensor $h_{ab}$ or line element 
$\mathrm{d}s^2=h_{ab}\mathrm{d}x^a\mathrm{d}x^b$ in spatial coordinates $x^a$%
, $a=1,2,3$. The space-time metric is 
\begin{equation}  \label{metric}
g^{\mu\nu}=h^{\mu\nu}-n^{\mu}n^{\nu}
\end{equation}
(where $h^{\mu\nu}=0$ if $\mu$ or $\nu$ is zero). The unit-normal term
simply adds the time component to the inverse spatial metric.

A second family of observers sees time change not along $n^{\mu}$ but along
a different timelike vector field $t^{\mu}$. We can always relate the two
notions of time direction by $t^{\mu}=Nn^{\mu}+N^{\mu}$ with a (lapse)
function $N$ and a spatial (shift) vector $N^{\mu}$ tangent to the first set
of spatial cross-sections ($n_{\mu}N^{\mu}=0$). If $t^{\mu}$ is normalized
as a timelike vector field, 
\[
\frac{t^{\mu}}{\sqrt{-||t||^2}}= \frac{n^{\mu}+N^{\mu}/N}{\sqrt{1-|\vec{N}%
/N|^2}}\,, 
\]
a comparison with 4-velocities $u^{\mu}=(1-|\vec{V}|)^{-1/2}(1,\vec{V})$ in
special relativity allows one to identify $\vec{N}/N$ as the 3-velocity of
the second family of observers with respect to the first.

The previous equation for the metric, (\ref{metric}), with $n^{\mu }$
expressed in terms of $t^{\mu }$, then provides the line element in ADM form 
\cite{ADM}  
\begin{equation}
\mathrm{d}s^{2}=-N^{2}\mathrm{d}t^{2}+h_{ab}(\mathrm{d}x^{a}+N^{a}\mathrm{d}%
t)(\mathrm{d}x^{b}+N^{b}\mathrm{d}t)
\end{equation}%
for coordinates such that $t^{\mu }\partial _{\mu }t=1$ and $t^{\mu
}\partial _{\mu }x^{a}=0$. In metric components, 
\begin{equation}
g_{00}=-N^{2}+h_{ab}N^{a}N^{b}\quad ,\quad g_{0a}=h_{ab}N^{b}\quad ,\quad
g_{ab}=h_{ab}\,.
\end{equation}%
For the inverse metric, we have components 
\begin{equation}
g^{00}=-\frac{1}{N^{2}},\quad g^{0a}=\frac{N^{a}}{N^{2}},\quad g^{ab}=h^{ab}-%
\frac{N^{a}N^{b}}{N^{2}}\,.  \label{ContrGQ}
\end{equation}%
Moreover, $\det g=-N\det h$.

\subsection{Stress-energy components}

Different families of observers assign different values as the
energy-momentum components of matter they measure. Our first family, moving
along the normal to its spatial cross-sections and often called ``Euclidean
observers,'' measures the energy density 
\begin{equation}  \label{rhoE}
\rho_{\mathrm{E}}=\frac{1}{\sqrt{\det h}} \frac{\delta H_{\mathrm{matter}}[N]%
}{\delta N}
\end{equation}
(the matter energy or Hamiltonian divided by the local volume), the energy
current 
\begin{equation}  \label{JE}
J^{\mathrm{E}}_a= \frac{1}{\sqrt{\det h}} \frac{\delta D_{\mathrm{matter}%
}[N^a]}{\delta N^a}
\end{equation}
with the momentum term $D_{\mathrm{matter}}$ of matter, and the spatial
stress 
\begin{equation}  \label{SE}
S^{\mathrm{E}}_{ab}= -\frac{2}{N\sqrt{\det h}} \frac{\delta H_{\mathrm{matter%
}}[N]}{\delta h^{ab}}\,,
\end{equation}
measuring how the matter energy reacts to spatial deformations that change
the metric. The trace of the spatial-stress tensor is proportional to the
pressure 
\begin{equation}
P_{\mathrm{E}}=-\frac{1}{N}\frac{\delta H_{\mathrm{matter}}[N]}{\delta \sqrt{%
\det h}}\,,
\end{equation}
the negative change of energy relative to the change of local volume.

In terms of the space-time stress-energy tensor $T_{\mu\nu}$, we identify (%
\ref{rhoE})--(\ref{SE}) as projections with respect to the unit normal and a
spatial frame $s_a^{\mu}$ of spacelike unit vectors with $n_{\mu}s_a^{\mu}=0$%
: 
\begin{equation}
\rho_{\mathrm{E}}= n^{\mu}n^{\nu}T_{\mu\nu}\quad,\quad J_a^{\mathrm{E}}=
n^{\mu}s_a^{\nu}T_{\mu\nu} \quad,\quad S_{ab}^{\mathrm{E}}=
s_a^{\mu}s_b^{\nu}T_{\mu\nu}\,.
\end{equation}
(These relationships also follow when one compares Einstein's equation with the canonical equations of motion of \cite{ADM}.)

A second, generic family of observers assigns energy-momentum components
according to the time direction $t^{\mu}$, for instance 
\begin{equation}
\rho=T_{00}=t^{\mu}t^{\nu}T_{\mu\nu}
\end{equation}
as the time component (or energy density measured by the new family of
observers). The relation between $\rho_{\mathrm{E}}$ and $\rho$ can be
obtained from $\rho_{\mathrm{E}}=n^{\mu}n^{\nu}T_{\mu\nu}$ together with the
relation between $n^{\mu}$ and $t^{\mu}$. For instance, 
\begin{eqnarray*}
\rho&=& t^{\mu}t^{\nu}T_{\mu\nu}= N^2 n^{\mu}n^{\nu} T_{\mu\nu}+
2Nn^{\mu}N^{\nu}T_{\mu\nu}+ N^{\mu}N^{\nu}T_{\mu\nu} \\
&=& N^2\rho_{\mathrm{E}}+2NN^aJ_a^{\mathrm{E}}+ N^aN^bS_{ab}^{\mathrm{E}}\,.
\end{eqnarray*}
Energy-momentum components therefore mix when the family of observers is
changed, just like local Lorentz transformations that change the
energy-momentum tensor.

We have expressed energy-momentum components as derivatives of the
Hamiltonian and momentum terms of matter by canonical metric components $N$, 
$N^a$ and $h_{ab}$. Equation (\ref{T}) expresses the full energy-momentum
tensor in terms of metric derivatives of the action. As one would expect,
these formulas are related. By Legendre transform, canonical expressions
appear in the action 
\begin{equation}  \label{action}
S_{\mathrm{matter}}= \int \mathrm{d}^4x \left(\dot{\phi}_Ip^I- N\mathcal{H}_{%
\mathrm{matter}}- N^a\mathcal{D}_a^{\mathrm{matter}}\right) \,,
\end{equation}
collecting all (not necessarily scalar) field components in $\phi^I$ with
momenta $p^I$. As written in this equation, the total matter Hamiltonian is
usually split in four components 
\begin{equation}
H_{\mathrm{matter}}[N,N^a]= \int\mathrm{d}^3x \left(N \mathcal{H}_{\mathrm{%
matter}}+ N^a \mathcal{D}_a^{\mathrm{matter}}\right)
\end{equation}
(integrated over a spatial cross-section) with four generators of shifts in
space-time, $\mathcal{H}_{\mathrm{matter}}$ in the time direction and $%
\mathcal{D}_a^{\mathrm{matter}}$ in spatial ones. Accordingly, $\mathcal{H}_{%
\mathrm{matter}}$ is the most important part for evolution, but for a given $%
t^{\mu}$ or $N$ and $N^a$, the actual evolution generator $H[N,N^a]$ is a
linear combination of all four components $\mathcal{H}$ and $\mathcal{D}_a$.
As we will see in the next subsection, the expressions of these components
must match delicately in order to ensure covariance, or independence of
one's choice of time. Quantum corrections in space-time evolution generators
must therefore be handled with extreme care. The same is true for
stress-energy components derived from them.

We derive stress-energy components in terms of derivatives of the
Hamiltonian by starting with the action 
\begin{eqnarray}  \label{varAction_g}
\delta S_{\mathrm{matter}}&=&-\frac{1}{2}\int\mathrm{d}^4x\sqrt{-\det g}%
T_{\mu\nu}\delta g^{\mu\nu}  \nonumber \\
&=&-\frac{1}{2}\int\mathrm{d}^4x N\sqrt{\det h}\left(T_{00}\delta
g^{00}+T_{a0}\delta g^{a0}+T_{0b}\delta g^{0b}+T_{ab}\delta g^{ab}\right)
\end{eqnarray}
varied by the metric \cite{KucharHypIII}. Inverse-metric variations are
related to variations of canonical metric components by 
\begin{eqnarray}  \label{G_vars}
\delta g^{00}&=& \frac{2\delta N}{N^3}  \nonumber \\
\delta g^{0a}&=& \frac{\delta N^a}{N^2}-\frac{2 N^a\delta N}{N^3} \\
\delta g^{ab}&=&\delta h^{ab}- \frac{1}{N^2}\left(N^a\delta N^b+N^b\delta
N^a-\frac{2N^aN^b}{N}\delta N\right)\,.  \nonumber
\end{eqnarray}
{}From 
\begin{eqnarray}  \label{varAction_q}
\delta S_{\mathrm{matter}} &=&-\int\mathrm{d}^4x \sqrt{\det h}\left(\frac{%
\delta N}{N^2}\left(T_{00}-2 N^a T_{a0}+N^a N^b T_{ab}\right)\right. \\
&& \qquad\qquad\qquad\qquad \left.+\frac{\delta N^a}{N}\left(T_{a0}-N^b
T_{ab}\right)+\frac{N}{2}\delta h^{ab}T_{ab}\right)  \nonumber
\end{eqnarray}
we read off $\delta S_{\mathrm{matter}}/\delta N$, $\delta S_{\mathrm{matter}%
}/\delta N^a$ and $\delta S_{\mathrm{matter}}/\delta h^{ab}$ as linear
combinations of stress-energy components. The inverted equations 
\begin{eqnarray}
T_{00} &=& -\frac{N}{\sqrt{\det h}} \left(N\frac{\delta S_{\mathrm{matter}}}{%
\delta N}+ 2N^a \frac{\delta S_{\mathrm{matter}}}{\delta N^a}+ 2\frac{N^aN^b%
}{N^2}\frac{\delta S_{\mathrm{matter}}}{\delta h^{ab}}\right)  \label{T00} \\
T_{0a} &=& - \frac{N}{\sqrt{\det h}} \left(\frac{\delta S_{\mathrm{matter}}}{%
\delta N^a}+ 2\frac{N^b}{N^2}\frac{\delta S_{\mathrm{matter}}}{\delta h^{ab}}%
\right) \\
T_{ab} &=& -\frac{2}{N\sqrt{\det h}} \frac{\delta S_{\mathrm{matter}}}{%
\delta h^{ab}}  \label{Tab}
\end{eqnarray}
show how energy-momentum terms follow from canonical metric derivatives,
expressing (\ref{T}) in terms of canonical metric components. Since $S_{%
\mathrm{matter}}$ in (\ref{action}) does not depend on time derivatives of
the metric, we can use $\delta S_{\mathrm{matter}}/\delta g^{ab}= -\delta H_{%
\mathrm{matter}}/\delta g^{ab}$ to obtain derivatives of the Hamiltonian.%
\footnote{%
This equation is somewhat subtle because it combines  4-dimensional
variations of space-time fields with 3-dimensional ones of  spatial fields,
and it may not be obvious that standard relations for  Legendre transforms
apply. Strictly speaking, we should write $\delta/\delta  g_{ab}(t,x)$ when
we vary the action, and $\delta/\delta g_{ab}(x)$ when we  vary the
Hamiltonian. Both functional derivatives are local expressions, so  that
there is no inconsistency in the equation. For the same reason,  $\dot{\phi}%
_I$ in (\ref{action}) must be considered independent of the  metric when the
action is varied, while the canonical  $\dot{\phi}_I=\{\phi_I,H_{\mathrm{%
matter}}[N,N^a]\}$ usually depends on the  metric. (For instance, $\dot{\phi}%
=Np_{\phi}/\sqrt{\det h}$ for a scalar  field.) Differentiating $S_{\mathrm{%
matter}}[\phi ,g]=\int \mathrm{d}t~\left(  \int \mathrm{d}^{3}x~p^I \lbrack
\phi ,g]\dot{\phi}_I-H_{\mathrm{matter}}[\pi  \lbrack \phi ,g],\phi
,g]\right) $ with respect to $g(t,x)$, one treats  $\phi_I $ and $g$ as the
independent variables (with any derivatives, e.g.  $\dot{\phi}_I$, being
functionally dependent on $\phi_I$ and $g$), as  opposed to $\phi_I$, $p^I$, 
$g$ (all assuming the gravitational momentum  does not appear). To arrive at
expressions for $\delta S_{\mathrm{matter}}[\phi  ,g]/\delta g(t,x)$ in
terms of $\delta H_{\mathrm{matter}}[\phi ,p ,g]/\delta  g(x)$ thus requires
use of the functional chain rule when considering  $H_{\mathrm{matter}}[p
\lbrack \phi ,g],\phi ,g]$; it is immediate to see that  its use neatly
cancels derivatives of $p^I \lbrack \phi ,g]\dot{\phi}_I$,  leaving only $%
-\delta H_{\mathrm{matter}}[p ,\phi ,g]/\delta g(x)$.} As the final step in
relating these different forms of energy-momentum components, we switch from 
$t^{\mu}$ to the normal. For instance, 
\begin{equation}  \label{rhoET}
T_{\mu\nu}n^{\mu}n^{\nu}= \frac{1}{N^2} (T_{00}-2 T_{0a}N^a+ T_{ab}N^aN^b)= -%
\frac{1}{\sqrt{\det h}} \frac{\delta S_{\mathrm{matter}}}{\delta N} = \rho_{%
\mathrm{E}}
\end{equation}
agrees with our previous formula for $\rho_{\mathrm{E}}$.

\subsection{Space-time symmetries}

The derivatives by $N$ and $N^a$ in (\ref{rhoE}) and (\ref{JE}) are by
canonical metric components, or equivalently by components of a space-time
vector field $t^{\mu}$ with respect to the normal $n^{\mu}$ and the spatial
cross-sections. They provide the energy and momentum densities, and
therefore exhibit the relation between energy-momentum and space-time
deformations. The same types of derivatives, applied to terms in the
gravitational Einstein--Hilbert action, show how space-time and its metric
change under deformations. If we change both matter fields and the
space-time metric according to some coordinate transformation, no
observables change --- there simply is no reference with respect to which
one could determine the change. These combined transformations must
therefore be symmetries of any physical theory, corresponding to general
covariance. The generators of these transformations are the derivatives of
the total action, obtained by adding gravitational and matter contributions,
by $N$ and $N^a$. We call the corresponding terms in the action 
\begin{equation}  \label{HNN}
H_{\mathrm{total}}[N,N^a]=-\int\mathrm{d}^3x \left(N\frac{\delta S_{\mathrm{%
total}}}{\delta N}+N^a\frac{\delta S_{\mathrm{total}}}{\delta N^a}\right)\,.
\end{equation}
Since they implement (coordinate) invariance of the complete system, their
values must be zero for all $N$ and $N^a$ when field equations are obeyed;
they function as constraints. There is an exact balance between
gravitational and matter energy and momentum.

Before equations of motion are solved for observables, we are dealing with
coordinate-dependent tensorial objects in space-time. They transform
non-trivially under coordinate changes or space-time deformations, which are
generated by the functionals $H_{\mathrm{total}}[N,N^a]$ depending on four
components of a space-time vector field. The algebra of these deformations
follows from Poisson brackets $\{H_{\mathrm{total}}[N,N^a],H_{\mathrm{total}%
}[M,M^a]\}$, computed by using a momentum $p^{ab}$ of $h_{ab}$ related to $%
\dot{h}_{ab}$, the derivative of $h_{ab}$ along $t^{\mu}$. A long
calculation, first completed by Dirac \cite{DiracHamGR} and interpreted
geometrically in \cite{Regained}, shows that 
\begin{eqnarray}
\{D_{\mathrm{total}}[M^a],D_{\mathrm{total}}[N^a]\}&=& -D_{\mathrm{total}%
}[N^b\partial_bM^a-M^b\partial_bN^a]  \label{DD} \\
\{H_{\mathrm{total}}[M],D_{\mathrm{total}}[N^a]\}&=& -H_{\mathrm{total}%
}[N^b\partial_bM]  \label{HD} \\
\{H_{\mathrm{total}}[M],H_{\mathrm{total}}[N]\}&=& D_{\mathrm{total}%
}[h^{ab}(M\partial_bN-N\partial_bM)]  \label{HH}
\end{eqnarray}
where $H_{\mathrm{total}}[N,N^a]=H_{\mathrm{total}}[N]+D_{\mathrm{total}%
}[N^a]$ according to (\ref{HNN}). This hypersurface-deformation algebra
encodes the classical structure of space-time: any theory with gauge
transformations obeying (\ref{DD})--(\ref{HH}) is generally covariant with
the classical space-time structure, and vice versa \cite{DiracHamGR}. (As we
will discuss in more detail below, some but not all of these relations are
obeyed seperately by matter terms, not just by the total constraints --- at
least in the absence of curvature coupling which we will assume for
simplicity.)

In canonical quantum gravity, one first turns the spatial metric $h_{ab}$
and its momentum into operators, which are then used to construct operators
for $H_{\mathrm{total}}[N]$ and $D_{\mathrm{total}}[N^a]$. Instead of
Poisson brackets one then computes commutators. These calculations are
complicated and remain incomplete, but currently there is a broad set of
mutually consistent results \cite%
{ConstraintAlgebra,LTBII,JR,ScalarHol,ScalarTensorHol,ModCollapse,ThreeDeform,TwoPlusOneDef,TwoPlusOneDef2,AnoFreeWeak}%
, obtained with different methods and in various models of loop quantum
gravity,\footnote{%
Similar deformations have been found using non-local matter  effects \cite%
{TensorialDeform}.} that show quantum corrections in the algebra, especially
(\ref{HH}). Instead of (\ref{HH}), one then has 
\begin{equation}
\{H_{\mathrm{total}}[M],H_{\mathrm{total}}[N]\}= D_{\mathrm{total}}[\beta
h^{ab}(M\partial_bN-N\partial_bM)]  \label{HHbeta}
\end{equation}
with a phase-space function $\beta$, while (\ref{DD}) and (\ref{HD}) remain
unchanged. Quantum space-time obeys different symmetries than classical
space-time, but the same number of generators is realized: no local
symmetries are broken and the theory is anomaly-free. (It remains unclear
whether the full theory of loop quantum gravity can be anomaly-free, but
there is encouraging evidence from the diverse set of models mentioned.)

A theory invariant under hypersurface deformations must be generally
covariant, showing the symmetry related to energy-momentum. Quantum
corrections in the deformation algebra, such as (\ref{HHbeta}), must then be
reflected in the energy-conservation law. To uncover implications of quantum
space-time for energy conservation, we first derive the classical relation
between (\ref{HH}) and (\ref{DT}).

\section{Energy conservation in canonical terms}

We rewrite the covariant conservation law (\ref{DT}) in terms of canonical
variables, making use of some of our expressions for energy-momentum in
terms of metric derivatives. We focus in our main calculations on the time
component of the law, $\nabla_{\mu}T^{\mu}{}_0=0$.

\subsection{Connection terms}

The connection terms look especially interesting because quantum corrections
in the conservation law could directly lead to quantum corrections of
differential geometry. However, both connection terms in (\ref{DT}) turn out
to be very generic.

The general 
\begin{equation}
\Gamma^{\rho}_{\mu\nu}= \frac{1}{2}g^{\rho\sigma} \left(\frac{\partial
g_{\sigma\mu}}{\partial x^{\nu}}+ \frac{\partial g_{\sigma\nu}}{\partial
x^{\mu}}- \frac{\partial g_{\mu\nu}}{\partial x^{\sigma}}\right)
\end{equation}
reduces to simpler versions with the form of indices required for $%
\nabla_{\mu}T^{\mu}{}_{\nu}=\partial_{\mu}T^{\mu}{}_{\nu}+
\Gamma^{\mu}_{\mu\rho}T^{\rho}{}_{\nu}- \Gamma^{\rho}_{\mu
\nu}T^{\mu}{}_{\rho}$. First, we have 
\begin{equation}
\Gamma^{\mu}_{\mu\rho}= -\frac{1}{2} \frac{\partial \log\det g}{\partial
x^{\rho}}= \frac{1}{N}\frac{\partial N}{\partial x^{\rho}}+ \frac{1}{\sqrt{%
\det h}}\frac{\partial\sqrt{\det h}}{\partial x^{\rho}}\,.
\end{equation}
The two terms $\partial_{\mu}T^{\mu}{}_{\nu}+
\Gamma^{\mu}_{\mu\rho}T^{\rho}{}_{\nu}$ can therefore be combined to 
\begin{equation}  \label{partialT}
\partial_{\mu}T^{\mu}{}_{\nu}+ \Gamma^{\mu}_{\mu\rho}T^{\rho}{}_{\nu} =\frac{%
\partial_{\mu}(N\sqrt{\det h} T^{\mu}{}_{\nu})}{N\sqrt{\det h}}\,.
\end{equation}
For $\nu=0$, we then deal with contributions to the Hamiltonian density $%
\mathcal{\ H}_{\mathrm{matter}}=N\sqrt{\det h}T^0{}_0$ and momentum
densities.

The last connection term in (\ref{DT}), using 
\[
g_{\rho\sigma}\Gamma^{\sigma}_{\mu\nu}= \frac{1}{2} \left(\frac{\partial
g_{\rho\mu}}{\partial x^{\nu}}+ \frac{\partial g_{\rho\nu}}{\partial x^{\mu}}%
- \frac{\partial g_{\mu\nu}}{\partial x^{\rho}}\right)\,, 
\]
takes the form 
\begin{equation}
\Gamma^{\rho}_{\mu\nu} T^{\mu}{}_{\rho}= \frac{1}{2} \frac{\partial
g_{\mu\rho}}{\partial x^{\nu}} T^{\mu\rho}\,,  \label{GammaT}
\end{equation}
the last two terms in $g_{\rho\sigma}\Gamma^{\sigma}_{\mu\nu}$ disappearing
by symmetry of $T^{\mu\rho}$. For $\nu=0$, we write 
\begin{equation} \label{GammaT}
\Gamma^{\rho}_{\mu 0} T^{\mu}{}_{\rho} =\frac{1}{2} \frac{\partial
g_{\mu\rho}}{\partial t} T^{\mu\rho}= -\frac{1}{N\sqrt{\det h}} \frac{%
\partial g_{\mu\rho}}{\partial t} \frac{\delta H_{\mathrm{matter}}}{\delta
g_{\mu\rho}}\,.
\end{equation}
Using the chain rule to transform from metric derivatives in space-time
tensor form to metric derivatives by canonical components $N$, $N^a$ and $%
h_{ab}$, we have 
\begin{equation}
\Gamma^{\rho}_{\mu 0} T^{\mu}{}_{\rho} =-\frac{1}{N\sqrt{\det h}} \left(%
\frac{\partial N}{\partial t}\frac{\delta H_{\mathrm{matter}}}{\delta N}+ 
\frac{\partial N^a}{\partial t}\frac{\delta H_{\mathrm{matter}}}{\delta N^a}%
+ \frac{\partial h_{ab}}{\partial t}\frac{\delta H_{\mathrm{matter}}}{\delta
h_{ab}}\right)\,.
\end{equation}
These terms can simply be combined with the time derivative of $T^{0}{}_0$
in (\ref{partialT}), as we will show now.

\subsection{Derivatives}

So far, the algebra of space-time deformations has played no role in the
terms of the conservation law. We now look at the partial derivatives in
more detail. Since time derivatives are canonically computed as Poisson
brackets with the total Hamiltonian $H_{\mathrm{total}}[N,N^a]=H_{\mathrm{%
grav}}[N,N^a]+H_{\mathrm{matter}}[N,N^a]$, 
\begin{equation}
\dot{f}=\frac{\partial f}{\partial t}= \mathcal{L}_{t^{\mu}} f= \{f,H_{%
\mathrm{total}}[N,N^a]\}
\end{equation}
for any phase-space function $f$, and the matter contribution $H_{\mathrm{%
matter}}[N,N^a]$ to the Hamiltonian is used to compute energy-momentum
components, the algebra should appear. However, there are several subtleties.

First, we raise an index in $T_{\mu\nu}$, writing 
\begin{equation}
T^{\mu}{}_{0}= g^{\mu\nu}T_{\nu 0}= (h^{\mu\nu}-n^{\mu}n^{\nu})T_{\nu 0}=
\left(h^{\mu a}+\frac{n^{\mu}}{N}N^a\right) T_{a0}- \frac{n^{\mu}}{N}
T_{00}\,.
\end{equation}
In particular, 
\begin{eqnarray}
T^0{}_0 &=& -\frac{1}{N^2} T_{00}+ \frac{N^a}{N} T_{a0}\,, \\
T^b{}_0 &=& \frac{N^b}{N}T_{00}+ \left(h^{ab}- \frac{N^aN^b}{N^2}\right)
T_{a0}\,.
\end{eqnarray}
These components, using (\ref{T00})--(\ref{Tab}), are related to
metric-derivatives of the Hamiltonian by 
\begin{eqnarray}
T^0{}_0 &=& -\frac{1}{N\sqrt{\det h}} \left(N\frac{\delta H_{\mathrm{matter}}%
}{\delta N}+ N^a\frac{\delta H_{\mathrm{matter}}}{\delta N^a}\right)= -\frac{%
1}{N\sqrt{\det h}} \mathcal{C}_{\mathrm{matter}}[N,N^a]\,, \\
T^b{}_0 &=& \frac{1}{N\sqrt{\det h}} \left(NN^b\frac{\delta H_{\mathrm{matter%
}}}{\delta N}+ \left(N^2h^{ab}+N^aN^b\right) \frac{\delta H_{\mathrm{matter}}%
}{\delta N^a}+ 2N^ch^{ba}\frac{\delta H_{\mathrm{matter}}}{\delta h^{ac}}%
\right)  \nonumber \\
&=& \frac{1}{N\sqrt{\det h}} \left(N^b \mathcal{C}_{\mathrm{matter}}[N,N^a]+
N^2h^{ab} \frac{\delta H_{\mathrm{matter}}}{\delta N^a}+ 2N^ch^{ba} \frac{%
\delta H_{\mathrm{matter}}}{\delta h^{ac}}\right)  \label{Tb0}
\end{eqnarray}
where 
\begin{equation}
\mathcal{C}_{\mathrm{matter}}[N,N^a]=N\frac{\delta H_{\mathrm{matter}}}{%
\delta N}+ N^a\frac{\delta H_{\mathrm{matter}}}{\delta N^a}= N\mathcal{H}_{%
\mathrm{matter}}+ N^a \mathcal{D}_a^{\mathrm{matter}}\,.
\end{equation}

Combining all terms, we write 
\begin{eqnarray}
N\sqrt{\det h}\nabla_{\mu}T^{\mu}{}_0 &=& \partial_0(N\sqrt{\det h}T^0{}_0)+
\partial_b(N\sqrt{\det h}T^b{}_0)  \nonumber \\
&&+ \frac{\partial N}{\partial t} \frac{\delta H_{\mathrm{matter}}}{\delta N}%
+ \frac{\partial N^a}{\partial t} \frac{\delta H_{\mathrm{matter}}}{\delta
N^a}+ \frac{\partial h_{ab}}{\partial t} \frac{\delta H_{\mathrm{matter}}}{%
\delta h_{ab}}  \nonumber \\
&=& -\partial_0\mathcal{C}_{\mathrm{matter}}[N,N^a]+ N^b\partial_b \mathcal{C%
}_{\mathrm{matter}}[N,N^a]+ (\partial_bN^b) \mathcal{C}_{\mathrm{matter}%
}[N,N^a]  \label{div1} \\
&& + \frac{\partial N}{\partial t} \mathcal{H}_{\mathrm{matter}}+ \frac{%
\partial N^a}{\partial t} \mathcal{D}_a^{\mathrm{matter}}+ \frac{\partial
h_{ab}}{\partial t} \frac{\delta H_{\mathrm{matter}}}{\delta h_{ab}}
\label{div2} \\
&& +\partial_b\left(N^2h^{ab}\frac{\delta H_{\mathrm{matter}}}{\delta N^a}+
2N^ch^{ba} \frac{\delta H_{\mathrm{matter}}}{\delta h^{ac}}\right)\,.
\label{partial}
\end{eqnarray}
We write the first two lines in the last expression as 
\begin{equation}  \label{div12}
(\ref{div1})+(\ref{div2})= -N\frac{\partial \mathcal{H}_{\mathrm{matter}}}{%
\partial t}- N^a \frac{\partial \mathcal{D}_a^{\mathrm{matter}}}{\partial t}
+ \mathcal{L}_{\vec{N}} \mathcal{C}_{\mathrm{matter}}[N,N^a] + \frac{%
\partial h_{ab}}{\partial t} \frac{\delta H_{\mathrm{matter}}}{\delta h_{ab}}
\end{equation}
with the Lie derivative $\mathcal{L}_{\vec{N}} \mathcal{C}_{\mathrm{matter}%
}[N,N^a]$ of the density-weighted $\mathcal{C}_{\mathrm{matter}}[N,N^a]$. We
will return to the terms in (\ref{partial}) after rewriting time derivatives
as Poisson brackets.

\subsection{Poisson brackets}

The time derivatives of $\mathcal{H}_{\mathrm{matter}}$ and $\mathcal{D}_a^{%
\mathrm{matter}}$ in (\ref{div12}) can be expressed as Poisson brackets with
the total constraint (\ref{HNN}), adding gravitational and matter
contributions; at this point the constraint algebra enters. We write 
\begin{equation}
\dot{\mathcal{H}}_{\mathrm{matter}}= \{\mathcal{H}_{\mathrm{matter}}, H_{%
\mathrm{total}}[N]+D_{\mathrm{total}}[N^a]\}= \{\mathcal{H}_{\mathrm{matter}%
}, H_{\mathrm{total}}[N]\}+ \mathcal{L}_{\vec{N}} \mathcal{H}_{\mathrm{matter%
}}
\end{equation}
noting that the total diffeomorphism constraint $D_{\mathrm{total}}[N^a]$
generates the Lie derivative along $N^a$. Similarly, 
\begin{equation}
\dot{\mathcal{D}}_a^{\mathrm{matter}}= \{\mathcal{D}_a^{\mathrm{matter}}, H_{%
\mathrm{total}}[N]+D_{\mathrm{total}}[N^a]\}= \{\mathcal{D}_a^{\mathrm{matter%
}}, H_{\mathrm{matter}}[N]\}+ \mathcal{L}_{\vec{N}} \mathcal{D}_a^{\mathrm{%
matter}}\,,
\end{equation}
where we are free to use $H_{\mathrm{matter}}[N]$ in the end because the
gravitational Hamiltonian does not depend on matter fields. Continuing with (%
\ref{div12}), we obtain 
\begin{eqnarray}
(\ref{div1})+(\ref{div2})&=& -N\{\mathcal{H}_{\mathrm{matter}}, H_{\mathrm{%
total}}[N]\}- N^a \{\mathcal{D}_a^{\mathrm{matter}}, H_{\mathrm{matter}%
}[N]\} +(N^a\partial_aN) \mathcal{H}_{\mathrm{matter}}  \nonumber \\
&&+ \frac{\partial h_{ab}}{\partial t} \frac{\delta H_{\mathrm{matter}}}{%
\delta h_{ab}} \,.  \label{Poisson}
\end{eqnarray}

The two Poisson brackets encountered here both contain one local constraint
and are therefore not directly given by (\ref{HH}) or (\ref{HD}). Moreover,
not all expressions in them refer to total constraints. Poisson brackets
with local functions can be obtained from those smeared with $N$ or $N^a$ by
functional derivatives. For instance, 
\begin{eqnarray}
\{\mathcal{H}_{\mathrm{matter}},H_{\mathrm{matter}}[N]\} &=& \frac{\delta
\{H_{\mathrm{matter}}[M],H_{\mathrm{matter}}[N]\}}{\delta M}= \frac{\delta
D^{\mathrm{matter}}[h^{ab}(M\partial_bN-N\partial_bM)]}{\delta M}  \nonumber
\\
&=& \mathcal{D}_a^{\mathrm{matter}}h^{ab}\partial_bN+ \partial_b(\mathcal{D}%
_a^{\mathrm{matter}} h^{ab}N)= 2\mathcal{D}_a^{\mathrm{matter}}D^aN+ND^a%
\mathcal{D}_a^{\mathrm{matter}}  \nonumber \\
&=& \frac{1}{N}D^a(N^2\mathcal{D}_a^{\mathrm{matter}})\,.  \label{HHlocal}
\end{eqnarray}
Here, we have assumed the matter Hamiltonian to be free of curvature
couplings, so that the matter terms obey the same Poisson relation (\ref{HH}%
) as the total constraints. Moreover, we have substituted covariant spatial
derivatives $D_a$ (with $D_ah_{bc}=0$) for partial ones. The Poisson bracket
required for (\ref{Poisson}), which has the total constraint, is then 
\begin{equation}  \label{Hmattertotal}
\{\mathcal{H}_{\mathrm{matter}}, H_{\mathrm{total}}[N]\}= \frac{1}{N}D^a(N^2%
\mathcal{D}_a^{\mathrm{matter}})+ \{h_{ab},H_{\mathrm{grav}}[N]\} \frac{%
\delta \mathcal{H}_{\mathrm{matter}}}{\delta h_{ab}}\,.
\end{equation}
The last term writes $\{\mathcal{H}_{\mathrm{matter}}, H_{\mathrm{grav}}[N]\}
$ with an explicit variation by the metric, the only function in $\mathcal{H}%
_{\mathrm{matter}}$ with a non-trivial flow generated by $H_{\mathrm{grav}%
}[N]$.

The Poisson bracket with a diffeomorphism constraint in (\ref{Poisson}) can
be rewritten as a Lie derivative of $H_{\mathrm{matter}}[N]$ provided that
we add a term for the Lie derivative of the metric. (Unlike the Hamiltonians
in the absence of derivative couplings, the diffeomorphism and Hamiltonian
contributions from matter do not obey the same Poisson bracket as the total
constraints.) We have 
\begin{eqnarray}
\{\mathcal{D}_a^{\mathrm{matter}},H_{\mathrm{matter}}[N]\} &=& \frac{\delta
\{D_{\mathrm{matter}}[N^a],H_{\mathrm{matter}}[N]\}}{\delta N^a}  \nonumber
\\
&=& \frac{\delta}{\delta N^a}\left(H_{\mathrm{matter}}[N^a\partial_aN]+ \int%
\mathrm{d}^3x (\mathcal{L}_{\vec{N}}h^{ab}) \frac{\delta H_{\mathrm{matter}%
}[N]}{\delta h^{ab}}\right)  \nonumber \\
&=& \mathcal{H}_{\mathrm{matter}}\partial_aN+ 2D^b\frac{\delta H_{\mathrm{%
matter}}[N]}{\delta h^{ab}}\,,
\end{eqnarray}
using $\mathcal{L}_{\vec{N}}h^{ab}= -2D^{(a}N^{b)}$.

\subsection{Cancellations}

We return to (\ref{partial}) and combine all lines in the final expression
of this equation. We first replace the partial derivative in the last line
by a spatial covariant derivative, which can be done without correction
terms because we are dealing with the divergence of a vector field of
density weight one: 
\begin{equation}
(\ref{partial})= D_b\left(N^2h^{ab}\frac{\delta H_{\mathrm{matter}}}{\delta
N^a}+ 2N^ch^{ba} \frac{\delta H_{\mathrm{matter}}}{\delta h^{ac}}\right)=
D_b\left(N^2h^{ab}\mathcal{D}_a^{\mathrm{matter}}+ 2N^ch^{ba} \frac{\delta
H_{\mathrm{matter}}}{\delta h^{ac}}\right)\,.  \label{Dterm}
\end{equation}
Adding this result to (\ref{Poisson}) and using our expressions of Poisson
brackets shows that all terms cancel upon taking the derivative for all
factors in (\ref{Dterm}), observing the following identity: 
\begin{equation}
\dot{h}_{ab}= \{h_{ab},H_{\mathrm{total}}[N]+D_{\mathrm{total}}[N^a]\}=
\{h_{ab},H_{\mathrm{total}}[N]\}+ \mathcal{L}_{\vec{N}}h_{ab}= \{h_{ab},H_{%
\mathrm{grav}}[N]\}- 2D_{(a}N_{b)}\,.
\end{equation}

\section{Deformed energy conservation}

With our detailed analysis of the conservation law in the canonical
formalism we can see how modifications of space-time structure according to
deformations (\ref{HHbeta}) might change the classical form. Rather
surprisingly, the connection terms in (\ref{DT}) are not affected because
they simply serve to rewrite partial derivatives of energy-momentum
components as derivatives of Hamiltonians suitable for canonical
formulations. For this rewriting to work with modified space-time
structures, the connection components used are not to be modified, still
obeying the classical relationship with the metric. As always, one could add a tensor to the connection and try to absorb modifications due to the deformation. A modified $\partial_0\mathcal{C}_{\mathrm{matter}}[N,N^a]$ in (\ref{div1}), for instance, could be absorbed by such a tensor if the connection term (\ref{GammaT}) that gives rise to $\partial N/\partial t$ in (\ref{div2}) is changed by adding $N(\beta-1)$ to it. However, the remaining connection components are not to be changed because they cancel undeformed Poisson brackets involving the diffeomorphism constraint. The required terms added to connection components then do not form a covariant space-time tensor, and algebra deformations cannot be absorbed by changing the connection in a covariant way.
(Unused connection
terms, for instance the contributions to $\Gamma^{\rho}_{\mu\nu}$ that
dropped out by symmetry in (\ref{GammaT}), might be affected, but our
present considerations have nothing to say about them.)

Algebraic deformations (\ref{HHbeta}) therefore could only affect the
derivative terms in the conservation law. However, this alternative option
is again difficult to formulate in terms of stress-energy components because
the algebra, especially (\ref{HHlocal}) which now reads $\frac{1}{N}%
D^a(\beta N^2\mathcal{D}_a^{\mathrm{matter}})$, plays a role only for the
cancellation of one term $N^2h^{ab}\delta H_{\mathrm{matter}}/\delta N^a$ in 
$T^b{}_0$ in (\ref{Tb0}), whose spatial derivative is taken in (\ref{partial}%
). One cannot take $\beta$ into account by modifying the coefficient of $%
T^b{}_0$ in the conservation law, nor can $\beta$ be absorbed elsewhere
(such as in $N$ or $D_a^{\mathrm{matter}}$) because this would be in
conflict with other relations crucial for the final cancellations of $N$%
-dependent terms.

It is not possible to account for (\ref{HHbeta}) by simple modifications of
connection or derivative terms in (\ref{DT}). The only, but radical,
conclusion we can draw is that the usual space-time tensor calculus used to
define, among other things, the energy-momentum tensor and its covariant
derivative, completely breaks down in quantum space-time described by (\ref%
{HHbeta}). In fact, even the relation between the local and integrated forms
of the conservation law would be unclear in the absence of classical
coordinates and space-time manifolds.

With a simple modification of the standard covariant conservation law
unavailable, we can \emph{define} energy conservation more generally as the
closure of the constraint algebra including matter terms. Classically, this
condition implies (\ref{DT}) and energy conservation, and it is still
available in quantum gravity. Deformations of space-time structure in
quantum gravity require us to elevate the usual close relationship between
energy conservation and space-time symmetries to a principle: Energy
conservation and general covariance are not just related; they are one and
the same notion. They only appear conceptually different in classical
physics because several different but equivalent formulations of this law
are available. In the most general form of quantum gravity, only the version
referring directly to a closed and anomaly-free constraint algebra is
possible. When this condition is met, there is a local symmetry generator
along each direction in space-time, and we are allowed to say that energy
and momentum are conserved.

Even in the presence of deformed space-time structures, canonical space-time
descriptions are available and show complete consistency of the theory in
terms of gauge invariance and conservation laws associated with this
important symmetry. But common space-time formulations are too narrow to
encompass the modifications required by some theories of quantum gravity.
Taking (\ref{HHbeta}) into account, the only possible derivations of
observables and predictions are canonical, coupling gravity to matter as
developed in \cite{ConstraintAlgebra,ScalarGaugeInv}. In particular,
consistent cosmological perturbation theory based on matter perturbations
alone is not possible in the presence of quantum-geometry corrections in
loop quantum gravity.

In this context, it is crucial to have full control on the \emph{off-shell}
constraint algebra, and not just on a subset realized on complete or partial
solution spaces to the constraints. Often, one tries to side-step
complicated anomaly issues of the quantum constraint algebra by fixing the
gauge classically, or choosing a specific time and deparameterizing before
quantization. The resulting equations can formally be made consistent, but
one can no longer check whether they belong to a closed algebra of
quantum-corrected constraints. Most often, one chooses specific gauge
fixings or times based on mathematical simplicity alone, making it highly
unlikely that consistent algebraic structures are realized. Given our
conclusions, such models violate not only covariance but also energy
conservation.

\section{Conclusions}

We have found that deformed constraint algebras prohibit the use of standard
differential geometry because quantum corrections cannot be absorbed in
connection components or other ingredients of the conservation law. Details
of the derivation show that this conclusion is very general and insensitive
to the precise form of gravitational dynamics. Dynamical changes of the
metric always appear in the canonical conservation law in explicit form, by
the term $(\partial h_{ab}/\partial t) \delta H/\delta h_{ab}$ in (\ref%
{Poisson}). There is no need to refer to the metric dynamics in terms of
field equations for $h_{ab}$ or the explicit gravitational contribution $H_{%
\mathrm{grav}}$ to the constraints.\footnote{%
The gravitational constraint  algebra is important for the contracted
Bianchi identity to hold, as shown  in \cite{KucharHypII}.} As a
consequence, classical or quantum matter can be formulated consistently on
any background space-time of classical type, that is with undeformed Poisson
brackets (\ref{HH}). This fact is, of course, well known and used in quantum
field theory on curved space-time, whose backgrounds are not required to
solve Einstein's equation. But in contrast to assumptions often made in the
literature on quantum cosmology, quantum fields on \emph{\ quantum space-time%
} are much more subtle. Space-time structures are modified by quantum
effects, and the resulting correction terms, as shown here, cannot simply be
absorbed in appropriate conservation laws of standard type.

In our considerations, we used effective constraints and Poisson brackets
rather than fully quantized fields and commutators. However, our conclusions
are so universal that they apply to any system with deformed space-time
structure. There may be additional quantum corrections if one goes to higher
orders in effective equations or to the full quantum theory. But since these
corrections amount to the standard ones of quantum field theory, which do
not modify the conservation law, they cannot undo the effects pointed out
here.

Our results have important consequences for physical evaluations of modern
quantum cosmology: Any modification found in \emph{homogeneous}
minisuperspace models, such as bounce solutions often studied in such
contexts, can be implemented straightforwardly. After all, the modified
minisuperspace dynamics would just function as a new, non-Einstein
background for matter fields. Also metric perturbations can easily be
included if the gauge is fixed or if one uses deparameterization with a
single choice of time, for in these cases the constraint algebra is
circumvented. However, one does so by solving classical equations or
constraints before quantization, and therefore does not produce a consistent
quantum theory of space-time.

The great challenge of quantum gravity is to find a consistent set of
quantum constraints that implements a closed off-shell constraint algebra.
(See also \cite{NPZRev}.) This can be done only when matter is combined with
gravity and the full constraint algebra is considered without gauge fixing
or other restrictions. In loop quantum gravity, it then turns out \cite%
{ConstraintAlgebra} that (\ref{HH}) must be deformed to something of the
form (\ref{HHbeta}), and our conclusions derived from the conservation law
apply. Not only differential geometry but also ordinary quantum field theory
on curved space-time then becomes inapplicable, for the latter makes use of
traditional conservation laws.

\section*{Acknowledgements}

We are grateful to Stanley Deser for discussions. This work was supported in part by NSF grant PHY0748336.


\end{document}